# Image Gravimetry: A New Remote Sensing Approach for Gravity Analysis in Geophysics

M. Kiani *

1. M. Kiani, Department of Geomatics, University of Tehran, mostafakiani@ut.ac.ir

**Abstract**

In this paper a new Geophysical gravimetry approach is presented, which is based on satellite imagery in remote sensing. The method uses a satellite image, together with a set of points in the image the gravity values of which are known. Template-based spheroidal spline method of interpolation is used to constitute a system of equations to find the values of gravity at other points in the image. A real case study is presented for the Qom region in Iran. Values of gravity are determined from the Landsat satellite image for this region, using 9 points in the image whose gravity values are known. Reference ellipsoid gravity values, which are based on coefficients derived from satellite gravimetry, are computed for this region, as well. Comparison between gravity values derived from Landsat image and those from reference ellipsoid shows that the standard deviation of the results is around 6.71 milli Gal, with the maximum of differences being 35 milli Gal.

## 1. Introduction

New methods of monitoring the earth's surface have always been demanded and Geophysicists, with their physical perspective on phenomena, have devised many methods that could better model and represent earth's interior, surface and exterior. Among the most important concepts in Geophysics and Geodesy is the gravity, which has been the center of attention for a long time. Gravity is of both theoretical and applied interest to the Geophysicists, and there are a number of methods by which this quantity is gathered, among which are the gravimeters (of both absolute and relative type), and satellite gravimetry, with the latter being the most important advance in the field of gravity in the last decades.

Ground gravimetry by gravimeters is expensive and requires advanced instruments, in addition to being dependent on the environment's conditions (see [1] for more details). Satellite gravimetry is also expensive and esoteric, while being accurate and having global coverage. Its precision varies in different parts of the world and thus, it is considered a drawback of this method (see [2] for more details).

With the mentioned limitations of the ground and satellite gravimetry methods, we are motivated to develop a new approach of gravimetry. This method is to have the following characteristics:
1. Low cost of implementation; meaning it should be possibly a byproduct of another material already at our disposal.
2. Local and global analysis; meaning it is more general than the local ground gravimetry and global satellite gravimetry.
3. Low dependence on known values; such that even with one known value this method can be implemented.

The present purpose is focused on developing this new method. The first and second characteristics are satisfied when we use satellite images, with their primary purpose being other than gravimetry. These images can be used for both the local and global applications. The third characteristics is satisfied when we use a discrete, linear interpolant whose structure is based on a template. For this reason and noting that the earth's geometrical shape is close to an oblate spheroid, we have used the theory of spheroidal biharmonic splines in [3]. We thence call this new method "image gravimetry".

The rest of this paper is organized as follows. In section 2, the theory of this new approach is explained. In section 3 the case study of the theoretical results in section 1 is presented for the Qom region in Iran. The conclusions are stated in section 4.

## 2. Theoretical aspects of the image gravimetry

In this section the theoretical aspects of the image gravimetry is expounded and the steps to implement this method are presented.

The integral role of a satellite image is evident in this method. A satellite image here is assumed to have '$m$' rows and '$n$' columns, resulting in a total number of '$mn$' pixels in the image.

To implement image gravimetry, one first needs to transfer the satellite image coordinate system to the ground. This first step is crucial in that the method for finding the gravity values is based on the ground coordinates in latitude and longitude format. Hence, this so-called georeferencing process is the fundamental step in the image gravimetry approach. The image coordinate system, denoted by $(x, y)$, is defined based on the row and column number. In fact, $x$ is the row number and $y$ the column number. The ground coordinate system, denoted by $(X, Y, Z)$, is defined based on the longitude, $\lambda$, and co-latitude, $\theta$, as the following

$$(X, Y, Z) = (a \sin\theta \cos\lambda, a \sin\theta \sin\lambda, a \sqrt{1 - e^2} \cos\theta), \qquad (1)$$

in which $a$ and $e^2$ are, respectively, the semi-major axis and eccentricity of the reference ellipsoid. Usually the WGS84 ellipsoid is used for this reason. Here we have adopted this convention; thus $a = 6378137\ m$, and $e^2 = 0.0069$.

Note that it is better, though not necessary, to transfer the image coordinates to the ground system by the minimum required points. This is to prevent the image distortion effect, which will result in a distorted rectangular area on the ground. If the minimum number of points are used for the transformation, one can appreciate the fact that the rectangular image corresponds to a rectangular region on the ground and thus, the satellite image resolution determines the ground coordinates of each pixel. If the resolution is 'q' meters, then the approximate increments in co-latitudal and longitudal directions can be determined as follows

1. Co-latitude increment: based on the spheroidal mean meridian radius ($\bar{M}$)

Note that $\bar{M}$ is the average of meridian radius of the known point(s). If $v$ points are used for the transformation, then $\bar{M}$ is defined as the following

$$\bar{M} = \frac{1}{v}\sum_{i=1}^{v} M_i, \qquad (2)$$

where

$$M_i = \frac{a(1-e^2)}{(1-e^2\cos^2\theta_i)^{\frac{3}{2}}}. \qquad (3)$$

Then the co-latitudal increment, $d\theta$, is calculated as

$$q = \bar{M}d\theta \rightarrow d\theta = \frac{q}{\bar{M}}. \qquad (4)$$

2. Longitude increment: based on the spheroidal mean prime vertical radius ($\overline{N\sin\theta}$)

Here, one should note that the $v$ points can be used to find $\bar{N}$ as follows

$$\overline{N\sin\theta} = \frac{1}{v}\sum_{i=1}^{v} N_i \sin\theta_i, \qquad (5)$$

in which

$$N_i = \frac{a}{(1-e^2\cos^2\theta_i)^{\frac{1}{2}}}. \qquad (6)$$

The longitudal increment is then calculated as the following

$$q = \overline{N\sin\theta}\,d\lambda \rightarrow d\lambda = \frac{q}{\overline{N\sin\theta}}. \qquad (7)$$

In the case that more than the required number of points are used for georeferencing, mathematical transformations are needed, since the corresponding region on the ground is not perfectly rectangular. Since image is a two-dimensional space and ground is a three-dimensional one, a 2D to 3D transformation is required. One can use different types of this transformation, including Affine and DLT (see [4]). In the following, the mathematical forms of these functions are represented. In these relations, $(x, y)$ are the image coordinates (respectively row and column number of image point in the image coordinate system), and $(X, Y, Z)$ are the ground coordinates.

1. Affine transformation

$$\begin{pmatrix} x \\ y \end{pmatrix} = \begin{pmatrix} a_0 + a_1 X + a_2 Y + a_3 Z \\ b_0 + b_1 X + b_2 Y + b_3 Z \end{pmatrix}. \tag{8}$$

2. DLT transformation

$$\begin{pmatrix} x \\ y \end{pmatrix} = \begin{pmatrix} \dfrac{c_0 + c_1 X + c_2 Y + c_3 Z}{1 + c_8 X + c_9 Y + c_{10} Z} \\ \dfrac{c_4 + c_5 X + c_6 Y + c_7 Z}{1 + c_8 X + c_9 Y + c_{10} Z} \end{pmatrix}. \tag{9}$$

In each of the transformations above, the unknown coefficients ($a_i, b_i$ $i = 0, \ldots, 3$ in the first case and $c_i$ $i = 0, \ldots, 10$ in the second case) have to be determined to have the relationship between the image and ground spaces. In the first case at least four points are needed to find the coefficients. In the second case this number is 6. In order to find the co-latitudal and longitudal increments, the distance between all pixels and their adjacent points are calculated and converted to the co-latitude and longitude. Then all these values are averaged to give the final value of increments. In mathematical representations we have

$$(x_i, y_j) \xrightarrow{transformation} (X_{ij}, Y_{ij}, Z_{ij}), \quad i = 1, \ldots, m, \quad j = 1, \ldots, n, \tag{10}$$

$$(x_{i+1}, y_{j+1}) \xrightarrow{transformation} (X_{(i+1)(j+1)}, Y_{(i+1)(j+1)}, Z_{(i+1)(j+1)}), i = 1, \ldots, m-1,$$
$$j = 1, \ldots, n-1. \tag{11}$$

Then the distance ($\Delta r$) between the points is calculted

$$\Delta r_{ij}^{(i+1)(j+1)} = \sqrt{(X_{ij} - X_{(i+1)(j+1)})^2 + (Y_{ij} - Y_{(i+1)(j+1)})^2 + (Z_{ij} - Z_{(i+1)(j+1)})^2}, \tag{12}$$

Now, according to the relations in, the co-latitudal and longitudal increments are calculated as the following

$$d\theta_{ij}^{(i+1)(j+1)} = \frac{\Delta r_{ij}^{(i+1)(j+1)}}{\overline{M}}, \tag{13}$$

$$d\lambda_{ij}^{(i+1)(j+1)} = \frac{\Delta r_{ij}^{(i+1)(j+1)}}{\overline{N \sin\theta}}, \tag{14}$$

where $\overline{M}$ and $\overline{N\sin\theta}$ are the average of $M$ and $N\sin\theta$ for the point-pair with indices $(i, j)$ and $(i+1, j+1)$. Finally, the co-latitudal and longitudal increments are calculated as the following

$$d\theta = \frac{1}{(m-1)(n-1)} \sum_{i=1}^{m-1}\sum_{j=1}^{n-1} d\theta_{ij}^{(i+1)(j+1)}, \tag{15}$$

$$d\lambda = \frac{1}{(m-1)(n-1)} \sum_{i=1}^{m-1}\sum_{j=1}^{n-1} d\lambda_{ij}^{(i+1)(j+1)}, \tag{16}$$

After calculating $d\theta$ and $d\lambda$, the (approximate) ground coordinates of each pixel is calculated as the following

$$(\theta_i, \lambda_i) = (\min(\theta) + (i-1)d\theta, \min(\lambda) + (j-1)d\lambda), \tag{17}$$

where $\min(\theta)$ and $\min(\lambda)$ are the co-latitude and longitude of the first pixel in the image, calculated by the relations in.

So far, the image coordinates of each pixel are transferred to the ground, and the co-latitudal and longitudal increments are derived. In the final step, a method of interpolation is used to find the gravity values of each pixel in the image. For this reason, we have used the theory of template-based spheroidal splines in [3]. In this theory, it is proved that the following equation holds for each point, for the value of a function like $W$ at that point

$$\sum_{l=-2}^{+2}\sum_{k=-2}^{+2} d_{l,k}\, W(\theta_{i+l}, \lambda_{j+k}) = 0, \tag{18}$$

in which the coefficients $d_{l,k}$ are as the following

$$d_{-2,-2} = 0, \tag{19}$$

$$d_{-2,-1} = 0, \tag{20}$$

$$d_{-2,0} = \frac{\left(2 + d\theta\frac{\cot(d\theta-\theta)}{\sqrt{1-e^2\sin^2\theta}}\right)\left(2 - d\theta\frac{\cot(\theta)}{\sqrt{1-e^2\sin^2\theta}}\right)}{4h_\theta^4}, \tag{21}$$

$$d_{-2,1} = 0, \tag{22}$$

$$d_{-2,2} = 0, \tag{23}$$

$$d_{-1,-2} = 0, \tag{24}$$

$$d_{-1,-1} = \frac{\left(2 - d\theta\frac{\cot(\theta)}{\sqrt{1-e^2\sin^2\theta}}\right)\left(\frac{\csc^2(d\theta-\theta)+\csc^2(\theta)}{1-e^2\sin^2\theta}\right)}{2d\theta^2 d\lambda^2}, \tag{25}$$

$$d_{-1,0} = \frac{\left(2 - d\theta\frac{\cot(\theta)}{\sqrt{1-e^2\sin^2\theta}}\right)\left(-\frac{4}{d\theta^2} - \frac{2(\csc^2(d\theta-\theta)+\csc^2\theta)}{1-e^2\sin^2\theta}\right)}{2d\theta^2}, \tag{26}$$

$$d_{-1,1} = \frac{\left(2 - d\theta\frac{\cot(\theta)}{\sqrt{1-e^2\sin^2\theta}}\right)\left(\frac{\csc^2(d\theta-\theta)+\csc^2(\theta)}{1-e^2\sin^2\theta}\right)}{2d\theta^2 d\lambda^2}, \tag{27}$$

$$d_{-1,2} = 0, \tag{28}$$

$$d_{0,-2} = \frac{\csc^4\theta}{(1-e^2\sin^2\theta)^2 d\lambda^4}, \tag{29}$$

$$d_{0,-1} = \frac{2\left(\frac{\csc^2\theta}{1-e^2\sin^2\theta}\right)\left(-\frac{2}{d\theta^2} - \frac{2\csc^2\theta}{1-e^2\sin^2\theta}\right)}{d\lambda^2}, \tag{30}$$

$$d_{0,0} = 2\frac{\csc^4\theta}{(1-e^2\sin^2\theta)^2 d\lambda^4} + \left(-\frac{2}{d\theta^2} - \frac{2\csc^2\theta}{1-e^2\sin^2\theta}\right)^2 + \frac{\left(2-d\theta\frac{\cot(d\theta-\theta)}{\sqrt{1-e^2\sin^2\theta}}\right)\left(2-d\theta\frac{\cot(\theta)}{\sqrt{1-e^2\sin^2\theta}}\right)}{4d\theta^4}$$

$$+ \frac{\left(2+d\theta\frac{\cot(d\theta-\theta)}{\sqrt{1-e^2\sin^2\theta}}\right)\left(2-d\theta\frac{\cot(\theta+d\theta)}{\sqrt{1-e^2\sin^2\theta}}\right)}{4d\theta^4}, \tag{31}$$

$$d_{0,-1} = \frac{2\left(\frac{\csc^2\theta}{1-e^2\sin^2\theta}\right)\left(-\frac{2}{d\theta^2} - \frac{2\csc^2\theta}{1-e^2\sin^2\theta}\right)}{d\lambda^2}, \tag{32}$$

$$d_{0,2} = \frac{\csc^4\theta}{(1-e^2\sin^2\theta)^2 d\lambda^4}, \tag{33}$$

$$d_{1,-2} = 0, \tag{34}$$

$$d_{1,-1} = \frac{\left(2+d\theta\frac{\cot(\theta)}{\sqrt{1-e^2\sin^2\theta}}\right)\left(\frac{\csc^2(d\theta+\theta)+\csc^2(\theta)}{1-e^2\sin^2\theta}\right)}{2d\theta^2 d\lambda^2}, \tag{35}$$

$$d_{1,0} = \frac{\left(2+d\theta\frac{\cot(\theta)}{\sqrt{1-e^2\sin^2\theta}}\right)\left(-\frac{4}{d\theta^2} - \frac{2(\csc^2(d\theta+\theta)+\csc^2\theta)}{1-e^2\sin^2\theta}\right)}{2d\theta^2}, \tag{36}$$

$$d_{1,1} = \frac{\left(2+d\theta\frac{\cot(\theta)}{\sqrt{1-e^2\sin^2\theta}}\right)\left(\frac{\csc^2(d\theta+\theta)+\csc^2(\theta)}{1-e^2\sin^2\theta}\right)}{2d\theta^2 d\lambda^2}, \tag{37}$$

$$d_{1,2} = 0, \tag{38}$$

$$d_{2,-2} = 0, \tag{39}$$

$$d_{2,-1} = 0, \tag{40}$$

$$d_{2,0} = \frac{\left(2+d\theta\frac{\cot(d\theta+\theta)}{\sqrt{1-e^2\sin^2\theta}}\right)\left(2+d\theta\frac{\cot(\theta)}{\sqrt{1-e^2\sin^2\theta}}\right)}{4d\theta^4}, \tag{41}$$

$$d_{2,1} = 0, \tag{42}$$

$$d_{2,2} = 0. \tag{43}$$

Note that in the equation (18), if the value of $W(\theta_{i+l}, \lambda_{j+k})$ is known, then the right-hand side is not zero, since the value of $-d_{l,k} W(\theta_{i+l}, \lambda_{j+k})$ is transferred to the right side. If, however, the value of $W(\theta_{i+l}, \lambda_{j+k})$ is unknown, it is derived using the overall $m \times n$ equations. In fact, each pixel gives one equation and one unknown value, if its value is not known. It is important to notice that at least one point must be known to derive the other values. Note also that one can use other methods, such as spherical interpolant in [5]. However, this is not pursued here.

The method to derive the unknown values $W(\theta_{i+l}, \lambda_{j+k})$ is the usual matrix inversion. Since 12 of the coefficients $d_{l,k}$, $l, k = -2, ..., +2$ are zero, the matrix of the coefficients, denoted hereafter by D, is sparse. It can be proved that the inversion is performed with high stability.

The mathematical representation of the equations to find the unknown values $W(\theta_{i+l}, \lambda_{j+k})$ is as the following

$$DX = Y, \quad (44)$$

in which $X$ is the matrix of all unknown values $W(\theta_{i+l}, \lambda_{j+k})$, without, of course the known values, and $Y$ is a column matrix with most of its elements equal to zero, except when a known point exists in equation (18), in which case the corresponding value in $Y$ is $-d_{l,k} W(\theta_{i+l}, \lambda_{j+k})$.

Equation (44) could be solved using the usual inversion process, as the following

$$X = D^{-1}Y. \quad (45)$$

However, since the number of pixels (and thus the number of unknowns) in the satellite images is usually very large, the simple inversion process may not be applicable, since the matrix $D$ may not be even constituted. For instance, in the application we present in the next section, the satellite image has 2048 rows and 2048 columns, meaning the overall $4194304 \times 4194304$ size for $D$. Of course, it is not possible to handle this huge system of equations easily. Hence, we use the linear algebra techniques to deal with this difficulty.

One method to invert a matrix is to partition it into four sub-matrices, each of which having less dimensions than the original matrix (see references such as [6] and [7]). According to this method, the matrix $D$ is partitioned in the following format

$$D = \begin{pmatrix} P & Q \\ R & S \end{pmatrix}. \quad (46)$$

This partitioning allows us to find both the matrix $D$ and its inverse more easily. The $D^{-1}$ in the equation (45), is calculated as follows

$$D^{-1} = \begin{pmatrix} (P - QS^{-1}R)^{-1} & -(P - QS^{-1}R)^{-1}QS^{-1} \\ -S^{-1}R(P - QS^{-1}R)^{-1} & S^{-1} + S^{-1}R(P - QS^{-1}R)^{-1}QS^{-1} \end{pmatrix}. \quad (47)$$

Since in the equation primarily the inverse of $S$ is computed, one can think of a way to use the method in an iterative algorithm. For this reason, we have devised the following algorithm

1. Choose the size of $S$ such that a given computer could not invert it. If this size is $J \times J$, then the sizes of $P$, $Q$, $R$ are, respectively, $f \times g$, $(m \times n - J) \times (m \times n - g)$, and $(m \times n - f) \times (m \times n - J)$, in which $f, g < m \times n$ are the arbitrary dimension of matrix P.

2. Repartition the matrix $S$ as the following

$$S = \begin{pmatrix} S_{1,1} & S_{1,2} \\ S_{1,3} & S_{1,4} \end{pmatrix}. \quad (48)$$

After this partitioning, we try to find the inverse of S as the following

$$S^{-1} = \begin{pmatrix} (S_{1,1} - S_{1,2}S_{1,4}^{-1}S_{1,3})^{-1} & -(S_{1,1} - S_{1,2}S_{1,4}^{-1}S_{1,3})^{-1}S_{1,2}S_{1,4}^{-1} \\ -S_{1,4}^{-1}S_{1,3}(S_{1,1} - S_{1,2}S_{1,4}^{-1}S_{1,3})^{-1} & S_{1,4}^{-1} + S_{1,4}^{-1}S_{1,3}(S_{1,1} - S_{1,2}S_{1,4}^{-1}S_{1,3})^{-1}S_{1,2}S_{1,4}^{-1} \end{pmatrix}. \quad (49)$$

If again the $S_{1,4}^{-1}$ is not found, the $S_{1,4}$ is repartitioned. This process can go on until the highest possible dimension is derived for the $S_{i,4}$ in the iteration number $i$, as the following

$$S_{i,4} = \begin{pmatrix} S_{i+1,1} & S_{i+1,2} \\ S_{i+1,3} & S_{i+1,4} \end{pmatrix}. \tag{50}$$

3. After calculating $S$, compute the inverse of the matrix $D$ in (47) and therefore, the one in (45).

After calculating the inverse of matrix $D$, the unknown values can be computed, using the relation (45).

A summary of what is done in the image gravimetry approach is given in the following diagram

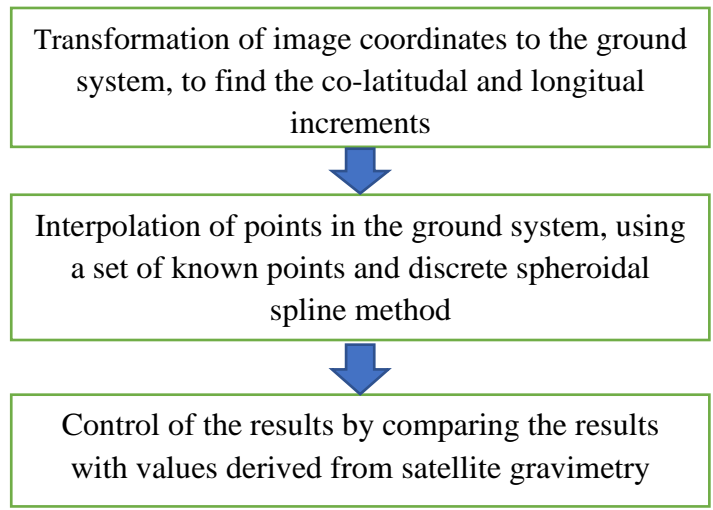

## 4. Application of the theoretical results: deriving gravity in the Qom region in Iran using the image gravimetry approach

In this section, a case study of the image gravimetry approach is presented for the Qom region in Iran. For the purpose of this paper, Landsat satellite images are used for this region. Figure 1 shows the region in the band numbers 1, 2, 3.

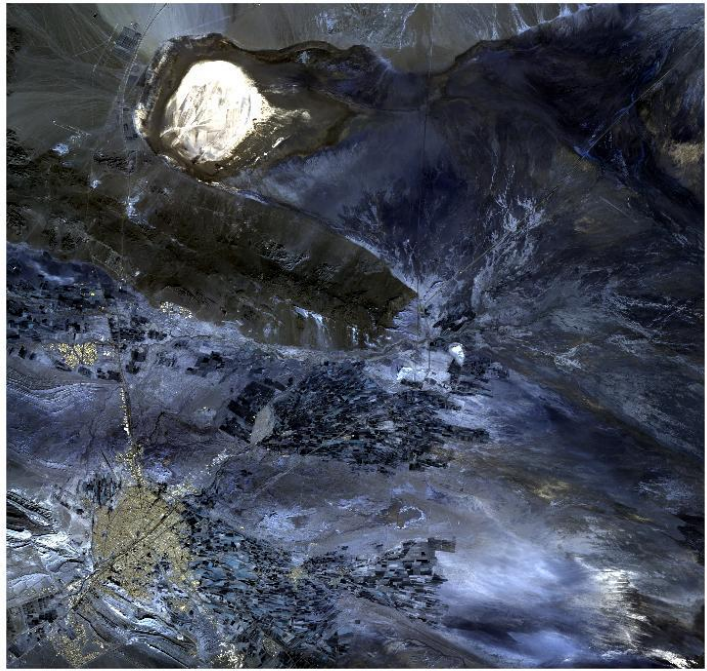

**Figure 1 :The region of case study (Qom, Iran)**

In order to calculate the gravity for the region, one first needs to choose some known points in the image. In the figure 2, the position of 9 known points is shown.

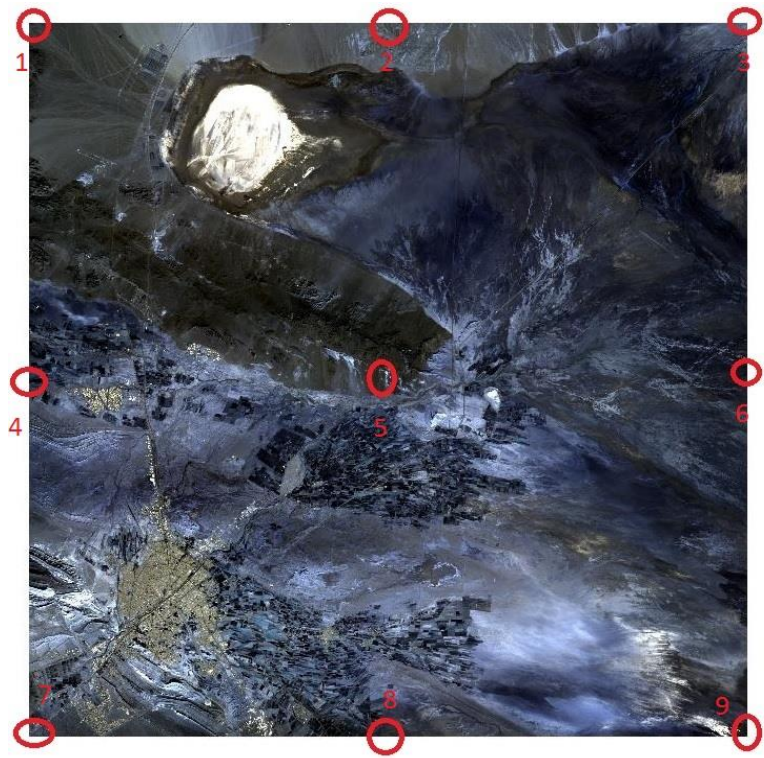

**Figure 2: Known points in the region of case study (Qom, Iran)**

In the following table the coordinates of these 9 points are mentioned.

Table 1: known points' coordinates and gravity values

| Point | $\theta°$ | $\lambda°$ | Gravity(mGal) |
|---|---|---|---|
| 1 | 55.875 | 49.142 | 981618.388 |
| 2 | 55.875 | 49.326 | 981609.953 |
| 3 | 55.875 | 49.511 | 981625.108 |
| 4 | 56.059 | 49.142 | 981582.901 |
| 5 | 56.059 | 49.326 | 981579.225 |
| 6 | 56.059 | 49.511 | 981589.102 |
| 7 | 56.243 | 49.142 | 981856.190 |
| 8 | 56.243 | 49.326 | 981858.843 |
| 9 | 56.243 | 49.511 | 981859.005 |

To implement the method of image gravimetry, the necessary elements $d\theta$ and $d\lambda$ are derived, using the relations in (15) and (16). Values of these parameters are

$$d\theta° = d\lambda° = 1.7996 \times 10^{-4}. \tag{51}$$

With these two parameters, the region can be described as

$$\theta° = [55.875, 56.243], \tag{52}$$
$$\lambda° = [49.142, 49.511]. \tag{53}$$

Figure 3 is the result of applying the image gravimetry approach to the Qom region.

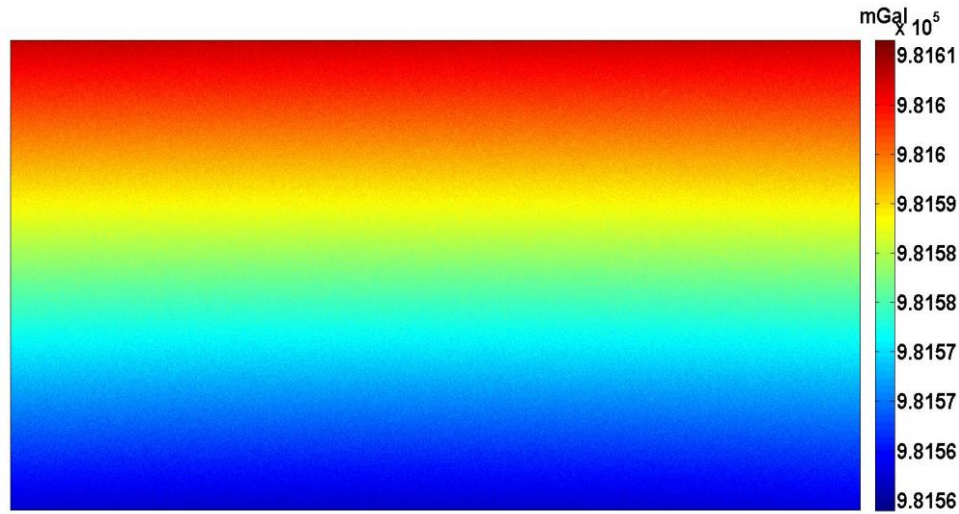

Figure 3: Gravity values in the Qom region, derived from the image gravimetry approach

### 4.1 Comparison with the satellite gravimetry approach

In order to analyze the differences between the satellite and image gravimetry approaches, the ellipsoidal gravity values, denoted by $g^e$, are computed in the region, using the relations below

$$g^e(\theta, \lambda) = \nabla W^e, \qquad (54)$$

where the ellipsoidal potential $W^e$ is defined as the following (see physical Geodesy references such [8])

$$W^e(\theta, \lambda) = \sum_{n=0}^{\infty} \sum_{m=0}^{n} P_{nm}(\cos\theta)(a_{nm} \cos m\lambda + b_{nm} \sin m\lambda) + \frac{1}{2} a^2 \omega^2 \sin^2\theta, \qquad (55)$$

in which $P_{nm}$ and $\omega$ denote, respectively, the Legendre polynomials of the first kind and the angular velocity of the earth's rotation. The gradient $\nabla$ in the equation is in the ellipsoidal coordinates. The coefficients $a_{nm}$ and $b_{nm}$ are derived from the satellite gravimetry approach. We have used coefficients in the EGM 2008 model. The following figure shows gravity values derived from this method

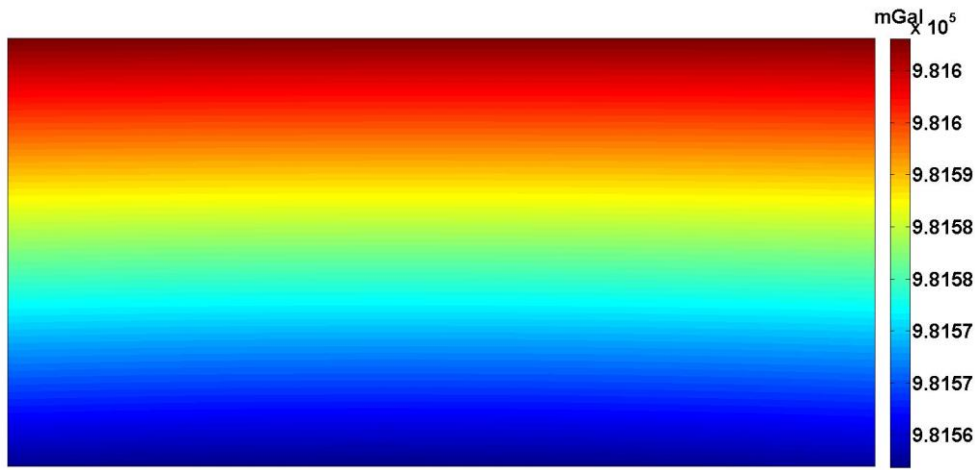

**Figure 4: Gravity values in the Qom region, derived from the satellite gravimetry approach**

To see the differences between image gravimetry and satellite gravimetry, values in figure 3 are subtracted from those in the figure 4. Figure 5 shows these difference values

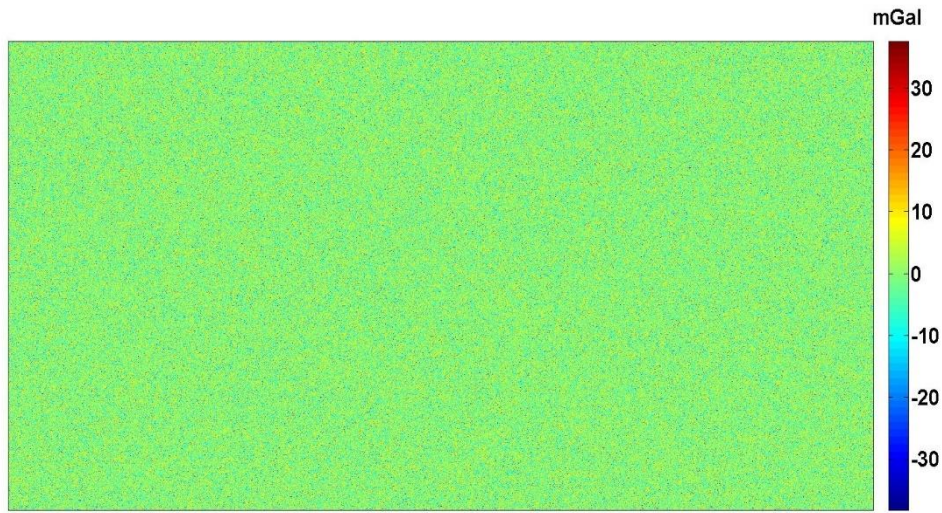

**Figure 5: Differences between image and satellite gravimetry approaches for the Qom region**

From the figure 5 it can be understood that the maximum of differences is around 35 mGal. The standard deviation of the results is also 6.71 mGal. To analyze the relative precision of the image gravimetry ($\delta g$), the average value of gravity in the region is derived from the figure 3. This value is $\bar{g} = 981582.179$ mGal. Hence

$$\delta g = \frac{35}{981582.179} = 3.566 \times 10^{-5}. \tag{56}$$

This relative precision can be considered quite good.

## 5. Conclusions

In this paper a new approach of gravimetry is presented. This method, called image gravimetry, is based on the satellite imagery in the field of Remote Sensing. Unlike the two other methods of gravimetry-ground and satellite-this method is more economical. It does not require the expensive instruments of ground gravimetry. It is also not very complex like the method of satellite gravimetry. It requires an image, a set of points in the image-whose spheroidal coordinates, $(\theta, \lambda)$, and gravity values, $g$, are known-and a discrete method of spheroidal interpolation. This is method of gravimetry is fast and stable, because of the optimized spheroidal interpolation. Comparison between the image and satellite gravimetry reveals that the maximum difference between these methods is 35 mGal, which is equivalent to relative precision of $3.566 \times 10^{-5}$. The power of this method is a motivation for many other works in the area of application of Remote Sensing to Geophysics. The introduction of image gravimetry can be a point of departure for future research on the improvement of this method.